\documentclass[11pt]{article}
\usepackage{epsfig}
\usepackage[font=small,labelfont=bf]{caption}
\usepackage[small,compact]{titlesec}

  \renewenvironment{thebibliography}[1]{%
    \begin{oldthebibliography}{#1}%
      \setlength{\parskip}{0ex}%
      \setlength{\itemsep}{0ex}%
  }%
  {%
    \end{oldthebibliography}%
  }

\def\beq{\begin{equation}}
\def\eeq#1{\label{#1}\end{equation}}
\def\eeqn{\end{equation}}

\def\beqa{\begin{eqnarray}}
\def\eeqa#1{\label{#1}\end{eqnarray}}
\def\eeqan{\end{eqnarray}}

\let\bar=\overbar

\def\Dslash{\not{\hbox{\kern-4pt $D$}}}
\def\dslash{\not{\hbox{\kern-2pt $\del$}}}

\def\msb{{\bar{\ssstyle M \kern -1pt S}}}

\newcommand{\pt}{p_{\rm T}}

\newcommand{\Dzero}{{\rm D^0}}
\newcommand{\Dstar}{{\rm D^{*+}}}
\newcommand{\Dplus}{{\rm D^+}}
\newcommand{\Dstrange}{{\rm D}_s}
\newcommand{\Gevc}{{\rm GeV}/{\it c}}

\def\Title#1{\begin{center} {\Large {\bf #1} } \end{center}}

\begin{document}

\Title{Heavy-flavour results in pp collisions at LHC with ALICE}

\bigskip\bigskip

\begin{raggedright}  

{\it Pietro Antonioli\index{Antonioli, P.} on behalf of the ALICE Collaboration\\
Istituto Nazionale di Fisica Nucleare\\
Sezione di Bologna\\
I-40126 Bologna, ITALY}
\bigskip\bigskip
\end{raggedright}

\vskip -1.0 cm

\section{Introduction}
Open heavy-flavour measurements at LHC are an important test of pQCD calculations based on the factorization approach in a new energy domain.  They provide also a baseline reference for heavy-ion collisions, where the heavy quarks produced in the early stages of the interactions are used to probe and characterize strongly interacting matter produced at high energy density and temperature.

The ALICE detector~\cite{ALICE-det} has good performance and specific detector characteristics to study open heavy flavour hadrons and quarkonia, at central and forward rapidities, thanks to its low momentum reach, particle identification capabilities and precise vertexing.

Open heavy flavour production is measured using semileptonic decays to electrons and muons or hadronic decays from charm mesons  ($\Dzero$, $\Dplus$, $\Dstar$ and D$_s$) at central rapidity. Recent results from measurements in pp collisions at different centre of mass energy (at $\sqrt s$~=~7 TeV and 2.76 TeV) are presented.

\section{Detector and data sample}
A detailed description of the ALICE detector can be found elsewhere~\cite{ALICE-det}. We highlight here only the key sub-detectors
employed in heavy-flavour analyses. In the barrel region, the Inner Tracking System (ITS) is the detector closest to the beam pipe and comprises three detector
layers, each using a different silicon technology. In the radial direction at $\eta$$<$0.9 the material budget is only 7.7\%~$\rm{X}_0$. 
This feature, coupled with a moderate (0.5~T) field in the barrel region, provides excellent coverage at low $\pt$. The vertex resolution
is below 100 $\mu$m at low multiplicity. At mid-rapidity ($|\eta|<0.9$) ALICE has powerful particle identification 
capabilities by means of its Time Projection Chamber (TPC), Transition Radiation Detector (TRD) and Time Of Flight (TOF) detectors, allowing 
the track by track identification of pions and kaons up to 2.5 $\Gevc$ momentum and electrons up to 8 $\Gevc$. Electron identification
is also provided by the electromagnetic calorimeter (EMCAL). Coupled with tracking reconstruction based on its large TPC, in the barrel
region ALICE identifies exclusive hadronic decay channels of charmed hadrons and semi-leptonic decays to electrons.
Muons are identified via the forward muon spectrometer in the pseudorapidity range 2.5$<$$\eta$$<$4. 

The results presented here are based on pp data samples collected at $\sqrt s$=7 TeV in 2010 and at $\sqrt s$=2.76 TeV in 2011 
with a minimum bias trigger based on ITS and V0 detectors (a scintillator array close to the beam pipe). The two data samples 
correspond, respectively, to an integrated luminosity $L_{\rm int}$=5 ${\rm nb}^{-1}$ and $L_{\rm int}$=1.1 ${\rm nb}^{-1}$.

\section{Hadronic decays of D mesons}
The measurement of charm production is carried out in ALICE through different channels ($\Dzero \rightarrow K^- \pi^+$, $\Dplus \rightarrow K^- \pi^+ \pi^-$, ${\rm D}^* (2010)^+ \rightarrow \Dzero \pi^+$ and $\Dstrange \rightarrow \phi \pi^+$)  and
results have been already published for three decay modes of D mesons, together with their charge conjugates at different center of mass energies~\cite{ALICE-D7TeV,ALICE-D276TeV}.
The ITS resolution allows for the identification of secondary vertices displaced few hundred $\mu$m from the primary vertex as the ones
associated to $\Dzero$ and $\Dplus$ mesons (their mean proper decay lenghts are ${\it c}\tau$ $\approx$ 123 and 312 $\mu$m respectively). 
The analysis strategy is based on the selection and reconstruction of secondary vertex topologies to reduce the large combinatorial background. The identification of charged kaons in the TPC and the TOF allow for further reduction of the background at low $\pt$.

\begin{figure}[!b]  
\begin{center}        
  \includegraphics[width=0.3\textwidth]{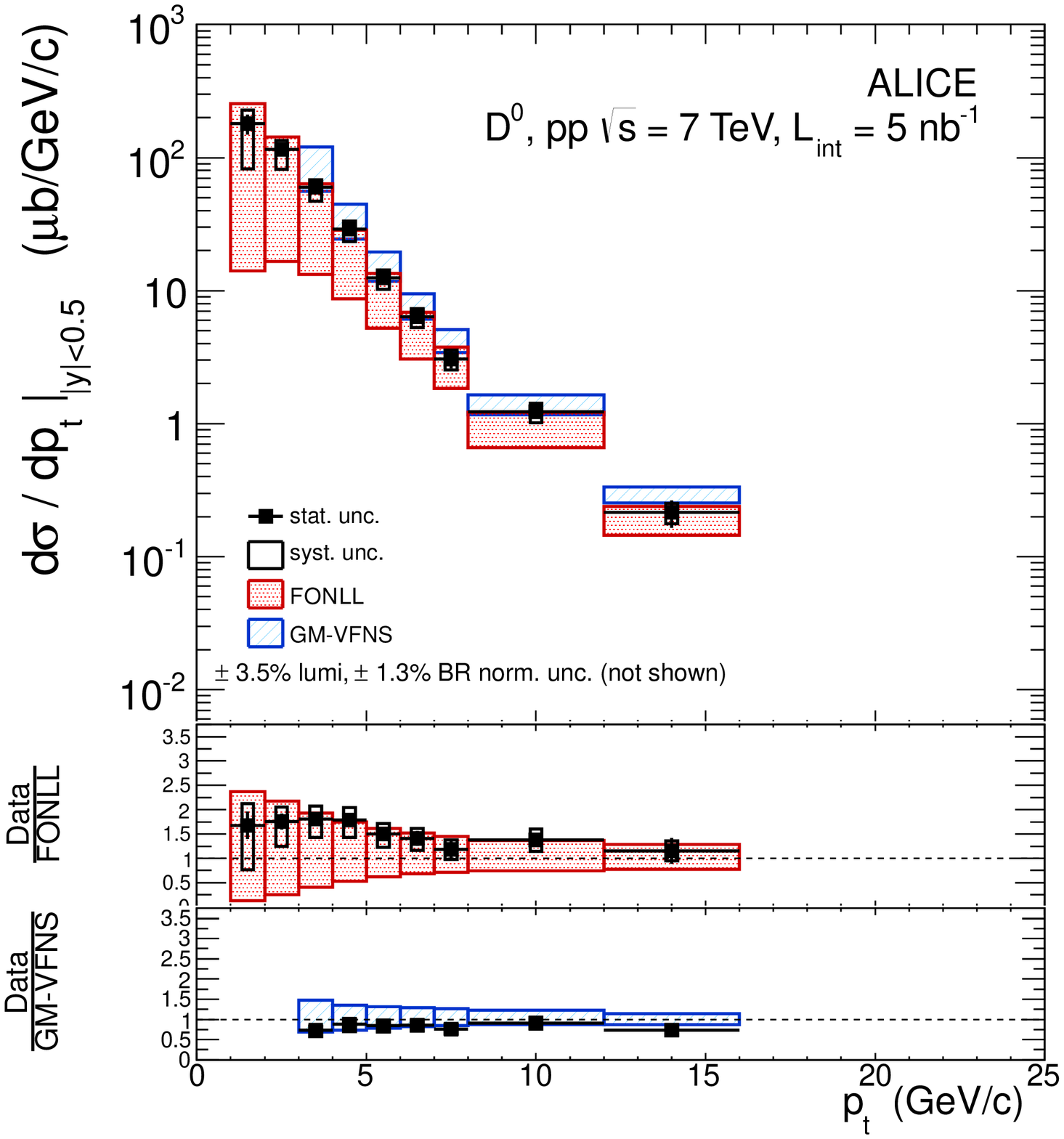}
  \includegraphics[width=0.3\textwidth]{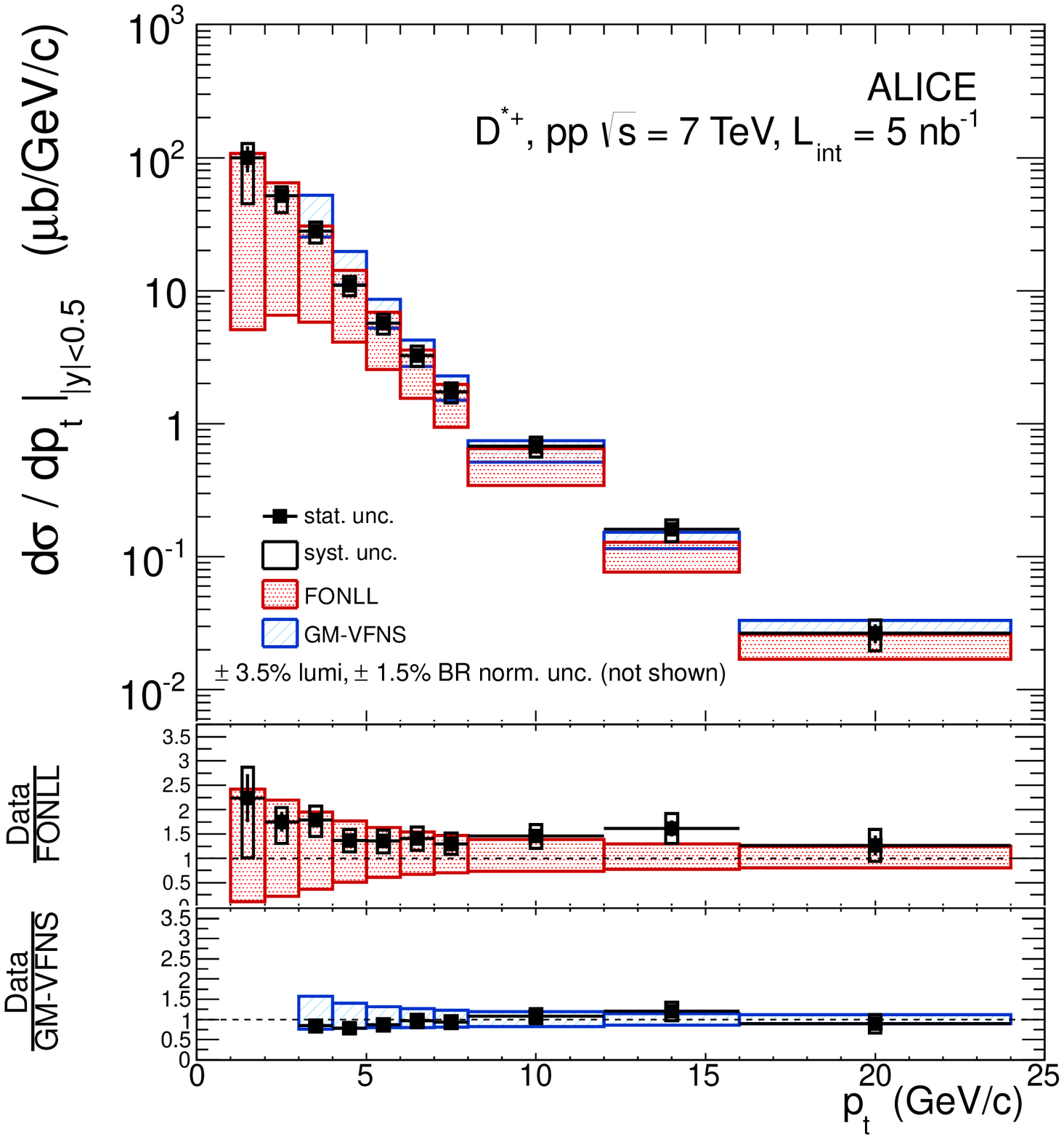}
  \includegraphics[width=0.3\textwidth]{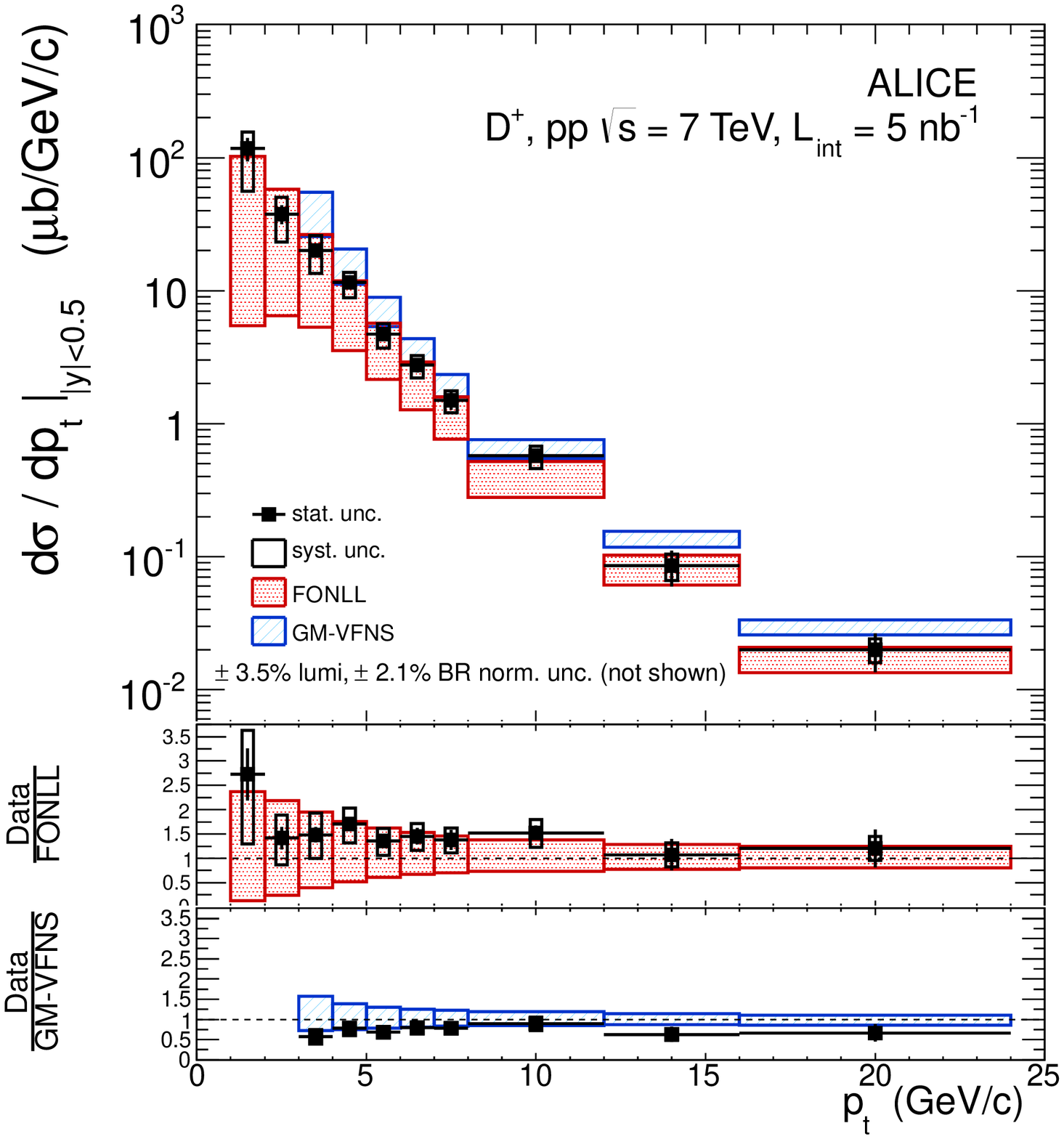}
  \caption{$\pt$-differential inclusive cross section for prompt $\Dzero$, $\Dstar$ and $\Dplus$ mesons for $|y|<0.5$ in pp collisions at $\sqrt{s}$~= 7 TeV~\cite{ALICE-D7TeV}, compared with FONLL~\cite{fonll} and GM-VFNS~\cite{gmvfns} theoretical calculations. The symbols are positioned horizontally at the centre of each $p_T$ interval. The normalization uncertainty is not shown (3.5\% from the minimum-bias cross section plus the branching ratio uncertainties).}
\label{fig:dmesons}
\end{center}
\end{figure}

The $\pt$-differential inclusive cross sections for prompt $\Dzero$, $\Dplus$ and $\Dstar$ are shown in Fig.~\ref{fig:dmesons}. 
The feed-down from B mesons decays (about 10-15\%) is subtracted using pQCD calculations.
The cross sections are well described by two pQCD-based predictions~\cite{fonll,gmvfns}. 

\begin{figure}[!t]
\begin{minipage}[t]{0.48\linewidth}
\centering
\includegraphics[width=0.95\textwidth]{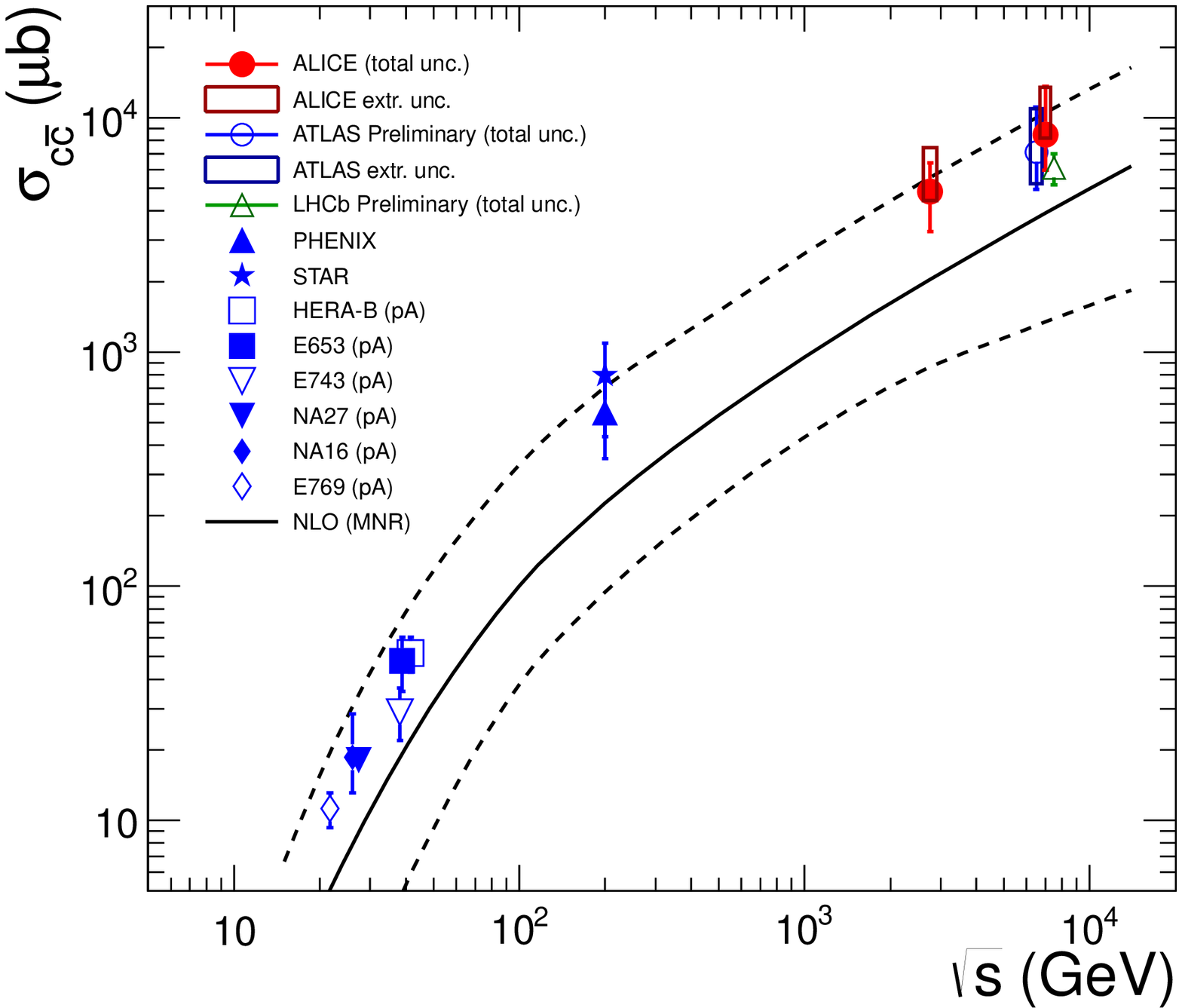}
\caption{Energy dependence of the total nucleon nucleon charm production cross section~\cite{ALICE-D276TeV}. 
The NLO MNR calculation (and its uncertainties) is represented by solid (dashed) lines.}
\label{fig:cctotal}
\end{minipage}
\begin{minipage}[t]{0.48\linewidth}
\centering
\includegraphics[width=0.95\textwidth]{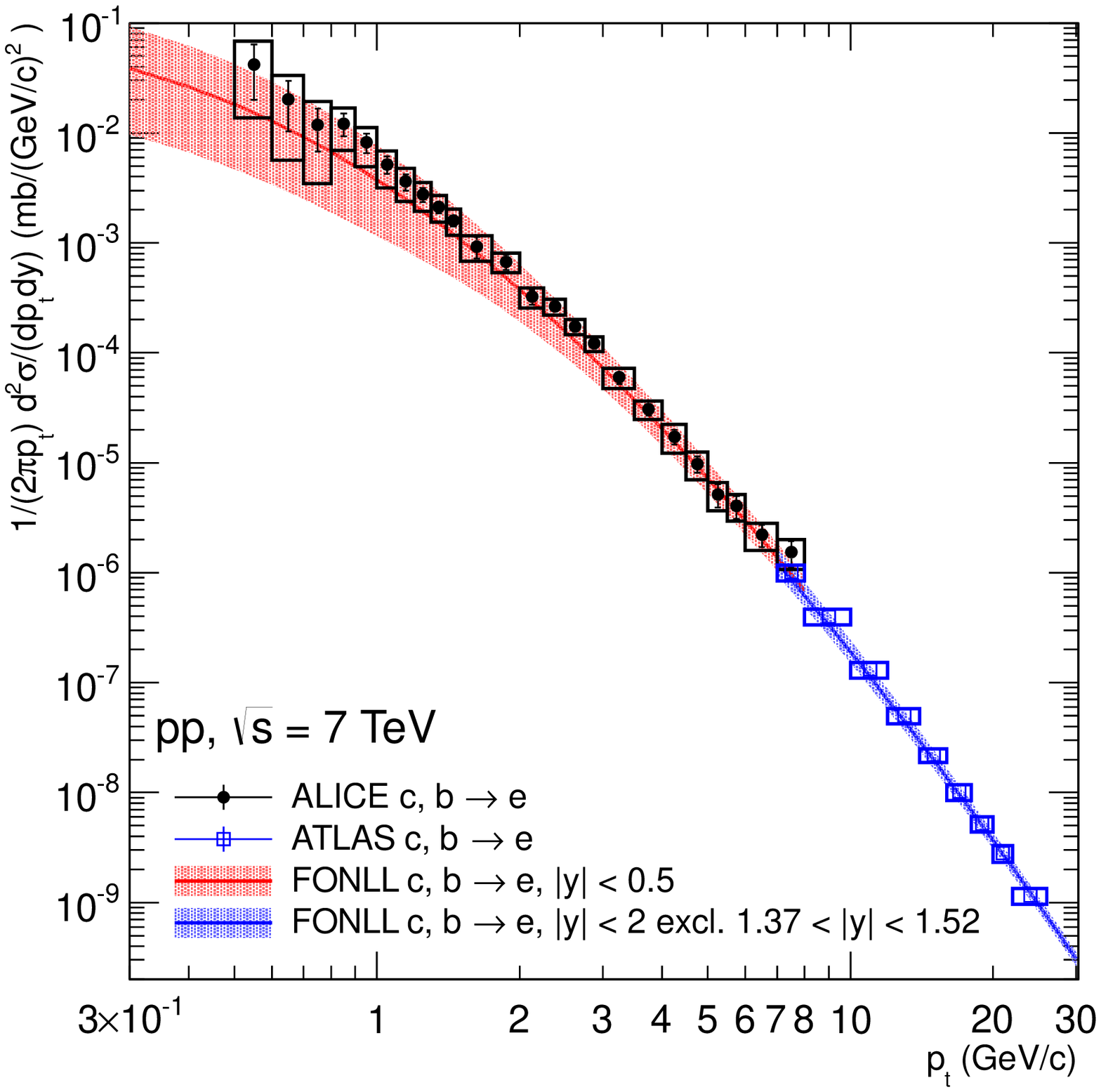}
\caption{Invariant differential production cross sections
of electrons from heavy-flavour decays measured by ALICE~\cite{ALICE-hfe} and
ATLAS~\cite{ATLAS-hfe} in pp collisions at $\sqrt{s} = 7$~TeV in different
rapidity intervals. FONLL pQCD calculations with the same rapidity
selections are shown for comparison.}
\label{fig:hfespectrum}
\end{minipage}
\end{figure}

The extrapolation of the $\rm{D}$ cross sections to the full phase space allows us to measure the total charm production cross section at LHC energies. A comparison with other measurements is shown in Fig.~\ref{fig:cctotal}, where ALICE results at $\sqrt s$=2.76 TeV~~\cite{ALICE-D276TeV} and 7 TeV ~\cite{ALICE-D7TeV} are shown. 
While a satisfactory agreement among LHC experiments is achieved, it may be noted that all points populate upper side of the theoretical predictions, which is based on NLO MNR calculations~\cite{mnr}.

A measurement~\cite{ALICE-jpsibb} of the total $\rm{b}\bar{\rm b}$ cross section production was 
also obtained. Other exclusive channels studied at ALICE include
$\Dstrange \rightarrow \phi \pi^+$~\cite{ALICE-Ds}. The study of the production rate of $D_s$ with respect to non-strange D mesons allows for
the investigation of the c fragmentation functions to strange and non-strange mesons.

\section{Semi-leptonic decays of heavy-flavour quarks}

Inclusive electron spectra coming from  heavy-flavour decays have been measured at LHC energies by
ATLAS in the 7$<$$\pt$$<$26 $\Gevc$ range~\cite{ATLAS-hfe}. The above mentioned features allow ALICE to make that measurement
at a much lower momentum.
Electrons are identified thanks to the energy deposit in the TPC and the timing in the TOF below 4~$\Gevc$. At higher momenta additional cuts
are applied making use of the TRD and the EMCAL detector information. The selection of tracks results in an almost pure sample 
of electrons with a remaining hadron contamination of less than 2\% over the full $\pt$ range. The heavy-flavour electron spectrum~\cite{ALICE-hfe} 
is then obtained on a statistical basis by subtracting a cocktail of background electrons from the inclusive electron spectrum. 
Systematic uncertainties on the measured electron spectrum due to the electron
cocktail amount to 10\%. Figure~\ref{fig:hfespectrum} shows the ALICE measurement, which includes
most of the total cross section, together with ATLAS data~\cite{ATLAS-hfe}, which extend the measurement at high $\pt$. Electrons from beauty decays 
are instead identified through displaced vertices~\cite{ALICE-be} exploiting the large ${\it c}\tau$ for B mesons ($\approx$~500~$\mu$m) or 
extracting the b-component from $\Delta\phi$ electron-hadron correlation studies.

\begin{figure}[!t]
\centering
\includegraphics[width=0.85\textwidth]{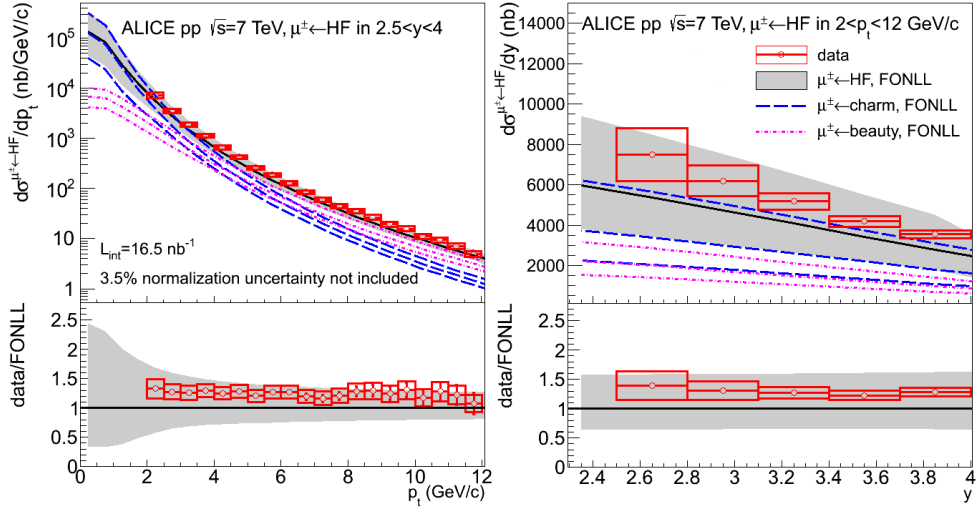}
\caption{$\pt$-differential (left) and y-differential (right) production cross section of muons from heavy-flavour decays~\cite{ALICE-singlemu}. In both panels, the error bars (empty boxes) represent the statistical uncertainties. The solid curves are FONLL calculations and the bands display the theoretical uncertainties. The FONLL calculations and systematic theoretical uncertainties for muons from charm and beauty decays are also shown.}
\label{fig:hfsinglemu}
\end{figure}

Single muons from heavy-flavour decays are studied at forward rapidity in ALICE using the forward
muon spectrometer. The subtraction of the background component from decay muons (muons from primary pion and kaon decays, mainly) is based on simulations.
Figure ~\ref{fig:hfsinglemu} shows the measured cross section  at $\sqrt s$=7 TeV~\cite{ALICE-singlemu}, as a function of $\pt$ and rapidity, compared to FONLL calculations. The measurement was also made at $\sqrt s$=2.76 TeV.

\vskip -0.5cm
\section{Conclusions}

Since the start of LHC operations, ALICE has produced a wide range of results related to heavy-flavour production in pp collisions
at $\sqrt s$=7 and 2.76 TeV center of mass energies. Within uncertainties, both FONLL and GM-VFNS describe well data.
Besides the interest to achieve a baseline reference for heavy-ion studies,
production cross sections for c and b quarks have been measured in a broad rapidity range and at very low momentum down to $\pt$=1-2~$\Gevc$, 
thus complementing measurements performed by other LHC detectors. Applying these analysis techniques
to forthcoming pPb data will allow for the investigation of possible nuclear modifications of the parton distribution functions.

\end{document}